\documentclass[a4paper,11pt]{article}
\usepackage{pos}
\usepackage{subcaption}
\usepackage{comment}
\usepackage{graphicx}
\usepackage{amsmath}

\usepackage{orcidlink}

\title{Likelihood and Deep Learning Analysis of the electron neutrino event sample at Intermediate Water Cherenkov Detector (IWCD) of the Hyper-Kamiokande experiment} \ShortTitle{IWCD electron neutrino event selection for Hyper-K experiment}

\author*[a]{T. Mondal \orcidlink{0000-0002-9445-1405}}
\author[b]{N. W. Prouse \orcidlink{0000-0003-1037-3081}}
\author[c]{P. de Perio \orcidlink{0000-0002-0741-4471}}
\author[d]{M. Hartz }
\author{D. Bose \orcidlink{0000-0003-1071-5854}$^{e}$ on behalf of the Hyper-Kamiokande Collaboration}

\affiliation[a]{Department of Physics, Indian Institute of Technology Kharagpur\\
Kharagpur, West Bengal 721302, India}
\affiliation[b]{Imperial College London, Department of Physics, London, United Kingdom}
\affiliation[c]{Kavli Institute for the Physics and Mathematics of the Universe (WPI), The University \\ of Tokyo Institutes for Advanced Study, University of Tokyo, Kashiwa, Chiba, Japan}
\affiliation[d]{TRIUMF, Vancouver, British Columbia, Canada}
\affiliation[e]{Department of Physics, Central University of Kashmir, Ganderbal, \\ Jammu $\&$ Kashmir 191131, India}

\emailAdd{mtanima14@gmail.com}

\abstract{Hyper-Kamiokande (Hyper-K) is a next-generation long baseline neutrino experiment. One of its primary physics goals is to measure neutrino oscillation parameters precisely, including the Dirac CP violating phase. As conventional $\nu_{\mu}$ beam generates from the J-PARC neutrino baseline contains only 1.5$\%$ of $\nu_{e}$ interaction of total, it is challenging to measure $\nu_{e}/\bar{\nu}_{e}$ scattering cross-section on nuclei. To reduce these systematic uncertainties, IWCD will be built to study neutrino interaction rates with higher precision. Simulated data comprise $\nu_{e}CC0\pi$ as the main signal with NC$\pi^{0}$ and $\nu_{\mu}CC$ are major background events. To reduce the backgrounds initially, a log-likelihood-based reconstruction algorithm to select candidate events was used. However, this method sometimes struggles to distinguish $\pi^{0}$ events properly from electron-like events. Thus, a Machine Learning-based framework has been developed and implemented to enhance the purity and efficiency of $\nu_{e}$ events.}

\FullConference{42nd International Conference on High Energy Physics (ICHEP2024)\\
18-24 July 2024\\
Prague, Czech Republic\\}

\begin{document}
\maketitle

\section{Introduction}
\vspace{-0.3cm}

Hyper-Kamiokande (Hyper-K) is a next-generation underground water Cherenkov detector, with a fiducial volume of eight times that of the existing Super-Kamiokande detector \citep{abe2018atmospheric} and an upgraded neutrino beam power of $\sim$1300 kW \citep{protocollaboration2018hyperkamiokande}. Its primary aim is to measure neutrino oscillation \citep{ashie2004evidence} by observing neutrinos from accelerator, atmospheric, solar, supernova and other astrophysical sources. One of the key physics goals of this long-baseline (L = 295 km) experiment is to investigate matter-antimatter asymmetry by measuring CP violation (CPV) in the neutrino sector \citep{Nunokawa_2008}.

A major challenge of this experiment lies in the precise measurement of electron neutrino cross-sections, which account for only $1.5\%$ of those created by conventional muon neutrino ($\nu_{\mu}$) beam. To reduce these systematic uncertainties, an Intermediate Water Cherenkov detector (IWCD) \citep{scott2016intermediate} will be positioned about $\sim 750$ m downstream from the J-PARC neutrino source. The IWCD comprises an inner detector (ID) radius of 400 cm, and height of 300 cm, representing a fiducial volume of $\sim600$ ton and equipped with $\sim500$ high-resolution multi-PMT (mPMT) \citep{protocollaboration2018hyperkamiokande} modules, each consisting of an array of 19 smaller 8 cm diameter PMTs. This provides a granularity comparable to that of Hyper-K’s far detector (FD). The detector volume of IWCD can move vertically within a 50 m tall pit, allowing the neutrino spectra to span $1^{\circ}-4^{\circ}$ off-axis angle \citep{protocollaboration2018hyperkamiokande} to study neutrino interaction rate at different peak energies with high accuracy. With its new detector technologies, IWCD will be capable of measuring the percent level electron neutrino (anti) ($\nu_{e}$ or  ${\bar{\nu}_{e}}$) contribution within $\nu_{\mu}$ beam, as well as measuring cross-section ratios ($\sigma _{\nu _{e}}/\sigma _{\nu _{\mu}}$, $\sigma _{{\bar{\nu}_{e}}}/\sigma _{{\bar{\nu}_{\mu}}}$). This will significantly improve the CPV sensitivity by reducing the associated systematic uncertainties.

To achieve these goals, advanced event reconstruction algorithms are essential for accurate particle identification (PID) and for the measurement of $\nu_{e}$ cross-sections. This work explores the $\nu_{e}$ event selection using both the traditional event reconstruction algorithm (fiTQun) and the advanced Machine Learning (ML) tool. The analysis presented here demonstrates how the ML techniques improve event selection, signal purity, and its efficiency, surpassing fiTQun.

\vspace{-0.3cm}
\section{IWCD Event reconstruction and Particle identification with fiTQun}
\vspace{-0.3cm}
Inside a water Cherenkov detector (WCD), when charged particles like electrons or muons travel faster than the speed of light in water, they polarize the surrounding atoms, causing the emission of cone-shaped blue light known as Cherenkov light. PMTs surrounding the detector volume record the number of these photons and their arrival time \citep{patterson2009extended}. By analysing the spatial and temporal distribution of the signal received by the hit PMTs, the event reconstruction algorithm determines, in part, particle type, momentum, energy and interaction location. Electrons produce broader, fuzzier Cherenkov rings due to multiple scattering and electromagnetic showering while they travel through the medium. Meanwhile, heavier muons undergo less scattering and typically produce a sharper Cherenkov ring.

\begin{figure}[ht]
    \centering
    % Text block in a minipage
    \begin{minipage}[b]{0.56\linewidth}
        \centering
        \includegraphics[width=\linewidth]{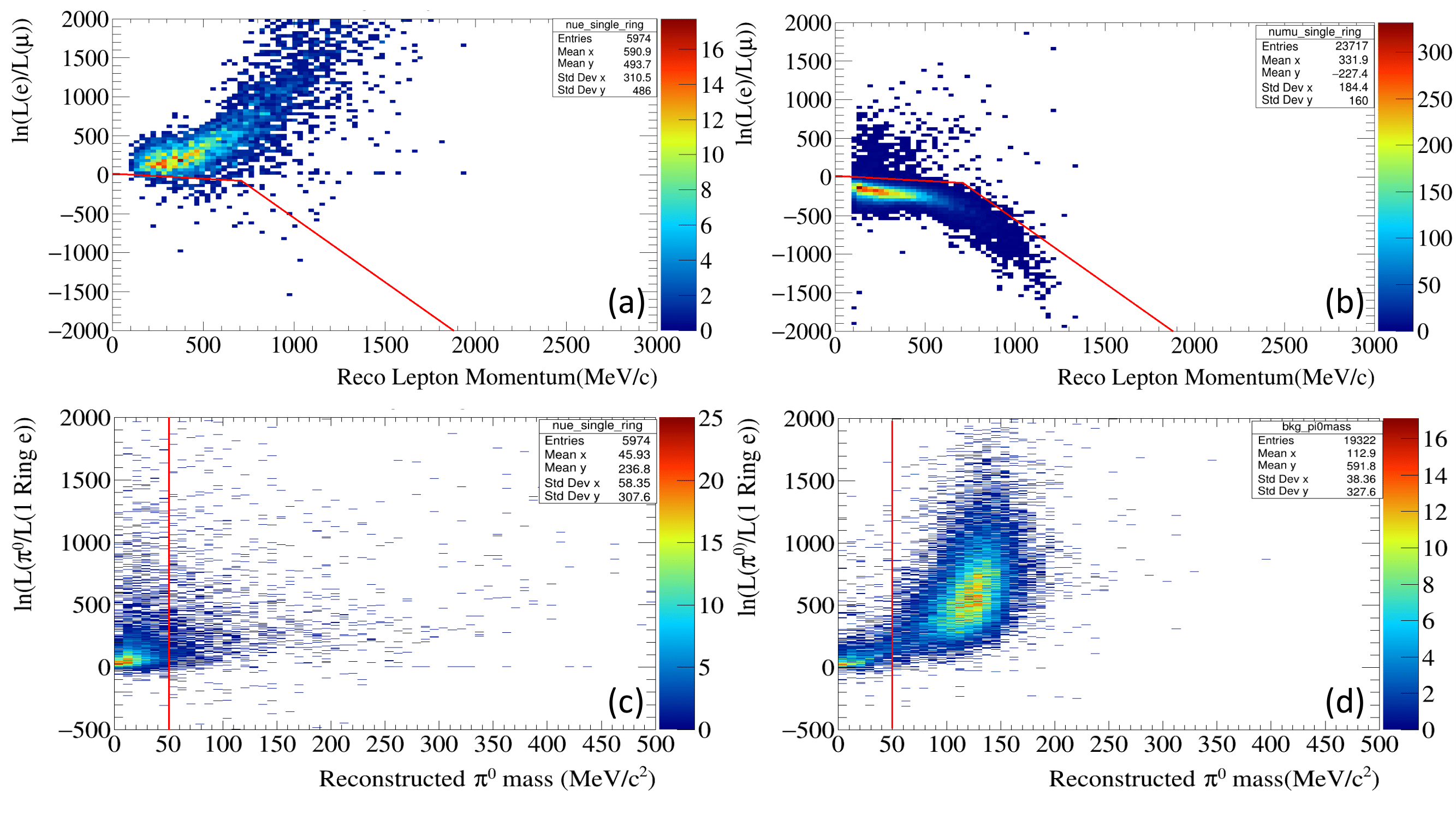}
        \caption{2D histograms depicting the distribution of $\nu_{e}CC0\pi$ signal in panels (a) and (c) and background events $\nu_{\mu}CC$ in panel (b), and $NC\pi^{0}$ in panel (d). In each panel the log-likelihood ratios are shown with respect to either the reconstructed lepton momentum or the reconstructed $\pi^{0}$ mass.}
        \label{fitqun_cuts}
    \end{minipage}
    \hfill
    % Figures in another minipage
    \begin{minipage}[b]{0.38\linewidth}
        \centering
        \includegraphics[width=\linewidth]{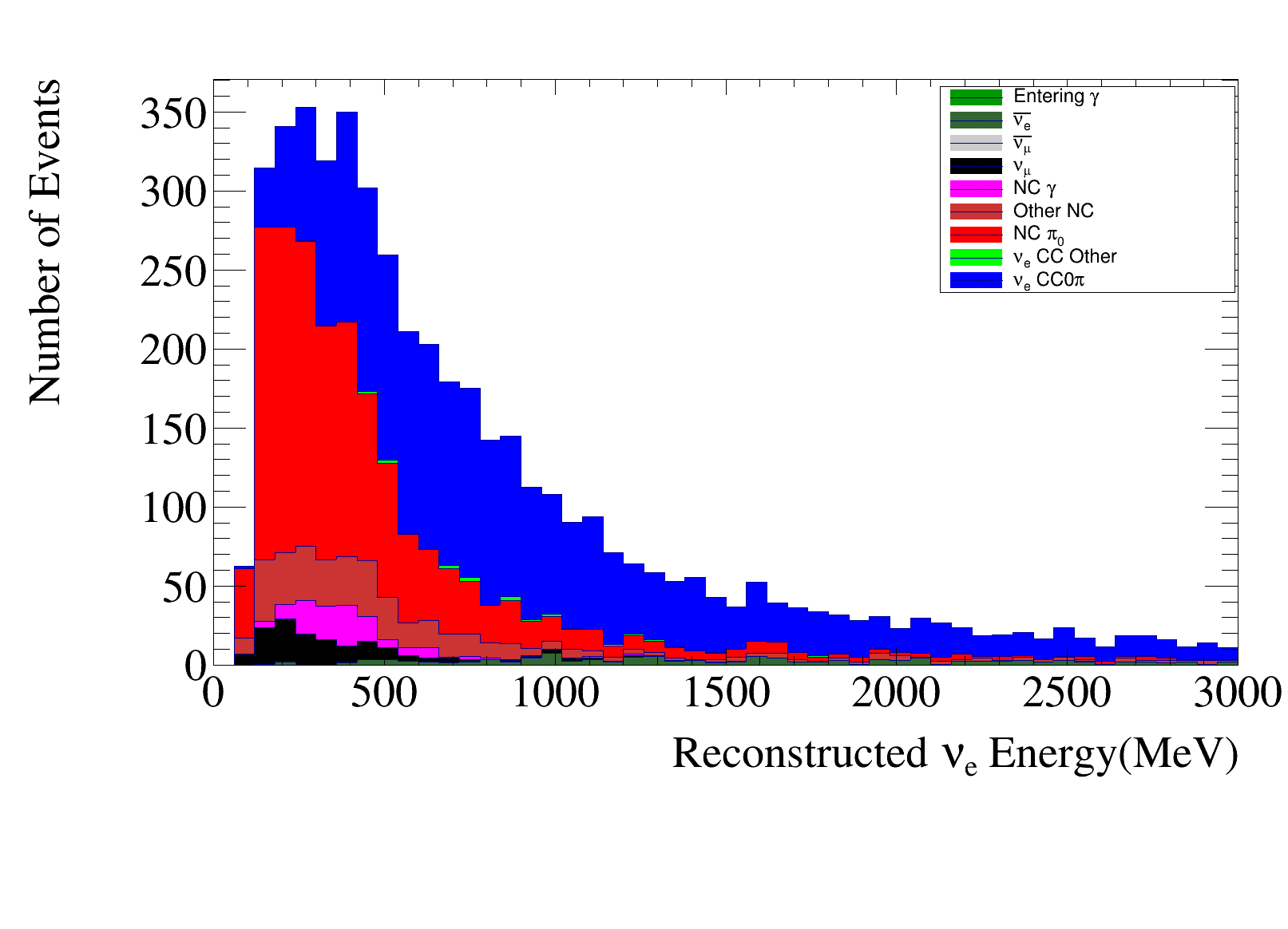}
        \caption{Interaction modes in the final selected $\nu_{e}$ events as a function reconstructed electron neutrino energy using fiTQun reconstruction algorithm.}
        \label{all_events_fiTQun} 
    \end{minipage}
    
\end{figure}

\subsection{IWCD event selection with fiTQun}

FiTQun \citep{patterson2009extended} is a likelihood-based event reconstruction algorithm used to reconstruct high-energy events inside WCD. It estimates the likelihood of the observed Cherenkov light pattern based on different particle hypotheses $\Gamma$ (such as $e^{-}$, $\mu^{-}$, or $\pi^{0}$) and reconstructs kinematic variables, including position, direction, and momentum, using the PMT's hit time and charge information. The likelihood ratios for different particle hypotheses are used to classify each event. For a given event hypothesis (e.g., electron or muon), the likelihood function \(L\) can be maximized by varying kinematic parameters $\theta$. The likelihood function is defined as follows \citep{missert2017improving,jiang2019atmospheric}:

\vspace{-0.5cm}
\begin{align} \label{equation_1}
    L(\Gamma, \theta)=\prod_{j}^{unhit}P_{j}\left ( unhit|\Gamma, \theta \right )\prod_{i}^{hit}\left \{ 1- P_{i}\left ( unhit|\Gamma, \theta\right )\right \}\times f_{q}\left ( q_{i}|\Gamma, \theta \right )f_{t}\left ( t_{i}|\Gamma, \theta \right )
\end{align}
\vspace{-0.55cm}

In this analysis, \(P_{j}(unhit|\Gamma, \theta)\) is the probability that a PMT does not register a hit for a given hypothesis \((\Gamma, \theta)\), while \(P_{i}(hit|\Gamma, \theta)\) is the probability of a PMT being hit. The reconstructed variables are defined by the kinematic parameters that maximize this likelihood. In the analysis, these parameters are Dwall (distance from the vertex to the nearest detector wall), Towall (distance to the nearest wall in the particle’s direction), kinetic energy, momentum and mass. They are crucial for distinguishing signal from background events. To accurately measure electron neutrino cross-sections, we have selected a sample enriched with electron neutrino interactions, where $\nu_{e}CC0\pi$ is the main signal and NC\(\pi^0\), and $\nu_{\mu}CC$ are backgrounds. Events are simulated at an off-axis angle of \(2.39^\circ\) with a total exposure of \(1 \times 10^{17}\) protons on target (POT) in FHC (Forward Horn Current) mode, where the beam is mainly composed of $\nu_{\mu}$. This analysis uses NEUT as the neutrino interaction generator, WCSim to simulate the detector response and fiTQun for the event reconstruction.

\textbf{Result ---} To significantly reduce background contamination and improve the PID accuracy, we define a fiducial volume (FV) with Dwall > 100 cm and Towall > 100 cm and select events with a reconstructed momentum of at least 100 MeV/c. Besides requiring single-ring events, additional cuts are applied to exclude events producing decay electrons and muons which penetrate the outer detector. To select $\nu_e $ candidates, relative log-likelihood cuts are used to remove $\nu_{\mu}CC$ and 
NC \(\pi^0\) events (see Figure \ref{fitqun_cuts}). The Figure of Merit ($FOM = \frac{S}{\sqrt{S+B}}$) is used to optimize the cut lines. In Figure \ref{fitqun_cuts}, the red lines represent the boundary of the cuts, below which events are classified as background. Figure \ref{all_events_fiTQun} shows the final selected events as a function of the reconstructed neutrino energy after applying all cuts and Table \ref{Comparison_table} summarizes these results. As one can see in Figure \ref{all_events_fiTQun}, $\nu_{\mu}CC$ are heavily suppressed, and the dominant remaining background in the final data sample is NC events. In the case of boosted $\pi^{0}$ decays, the two $\gamma$s can either overlap or one of them can carry most of the energy causing fiTQun to reconstruct the $\pi^{0}$ as a single ring electron-like event. This analysis achieved a $\nu_{e}CC0\pi$ purity of $51.1\%$ with an efficiency of $69.5\%$ (see Table \ref{Comparison_table}). FiTQun’s detailed and computationally complex likelihood allows it to process at most 1 event per minute. Improvements in accuracy require more complex likelihoods with fewer simplifying assumptions, further increasing the computational demand beyond a realistically feasible level.

\vspace{-0.4cm}
\section{Machine Learning (ML) for Electron Neutrino Event Selection}
\vspace{-0.4cm}

ML techniques, particularly Convolutional Neural Networks (CNNs) \citep{lecun2015deep}, offer significant potential for extracting information from complex images. These can contribute to high-accuracy particle identification \citep{Prouse:2021gr}. They also allow for reconstructing kinetic information (momentum, direction and interaction vertex) and for distinguishing between single and multi-ring events. In our analysis, we utilized a ResNet-18 architecture \citep{he2016deep}, which is an 18-layer CNN, within the WatChMaL \citep{WatChMaL} framework, to classify and reconstruct events inside both the IWCD and the Hyper-K far detector \citep{prouse2023machine}.

\vspace{-0.13cm}
\begin{figure}[ht!]
    \centering
    
    % Text block in a minipage
    \begin{minipage}[b]{0.62 \linewidth}
        This configuration offers faster training times compared to other deep neural networks. Initially, we trained our Resnet-18 model for 20 epochs on four-particle classes ($e^{-}$, $\mu^{-}$, $\gamma$, $\pi^{0}$) with 3 million of events each. The kinetic energy of each particle is uniformly distributed up to 1 GeV above the Cherenkov threshold. Figure \ref{gamma_rejection} compares the performance of fiTQun and ResNet-18 models from the ROC (Receiver Operating Characteristic) curve for $\gamma$ background rejection against electron signal efficiency. The ResNet model, with an AUC (Area Under the Curve) of 0.7183, significantly outperforms fiTQun (AUC of 0.5418) in separating electron signals from $\gamma$ backgrounds.
         
    \end{minipage}
    \hfill
    % Single figure in another minipage
    \begin{minipage}[b]{0.35\linewidth}
        \centering
        \includegraphics[width=\linewidth]{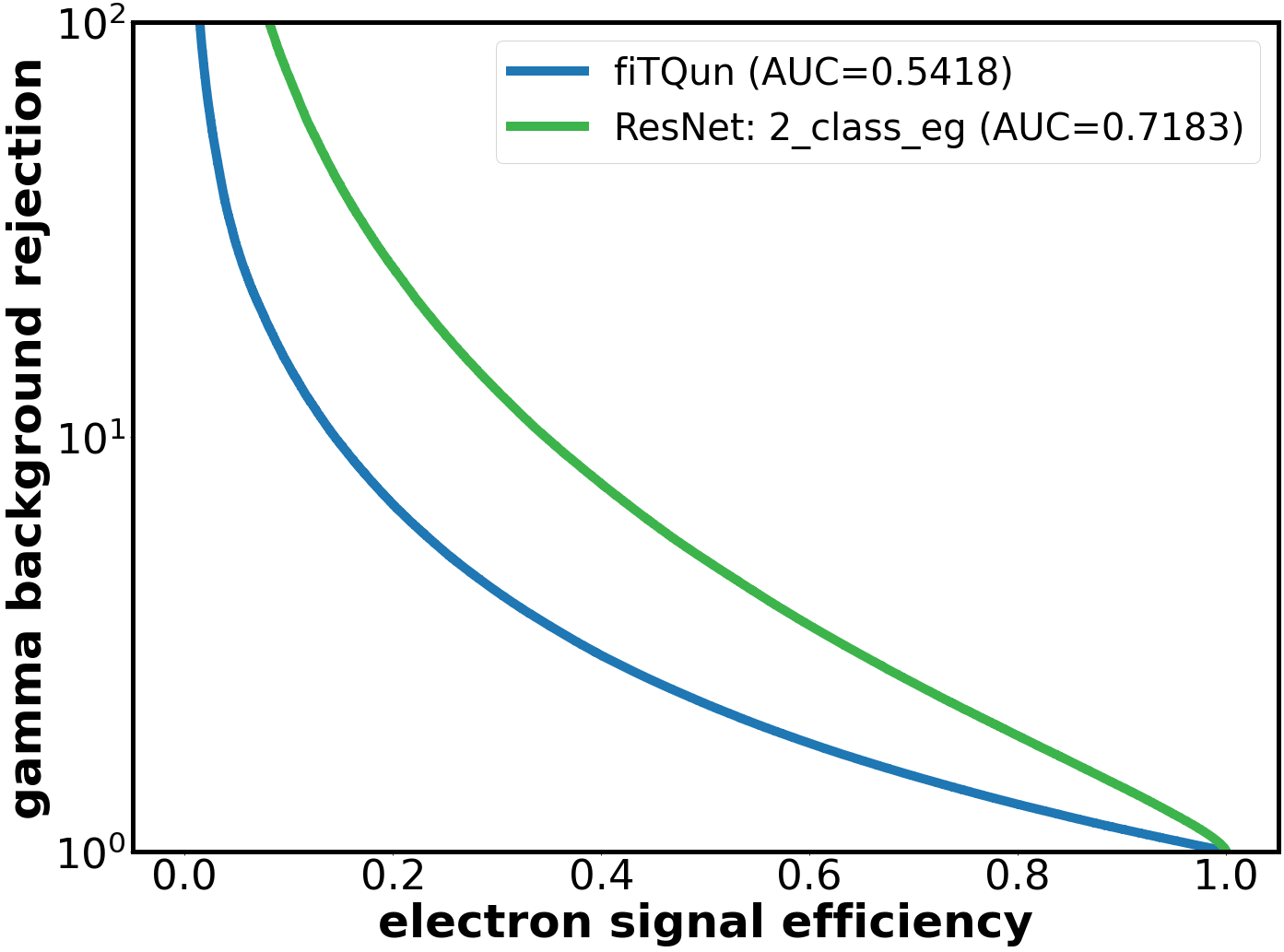}
        \caption{Comparison between fiTQun and ML performance in $\gamma$/electron separation.}
        \label{gamma_rejection}
    \end{minipage}
    
\end{figure}

\begin{figure}[ht]
    \centering
    % Text block in a minipage
    % Figures in another minipage
    \begin{minipage}[b]{0.56\linewidth}
        \centering
        \includegraphics[width=\linewidth]{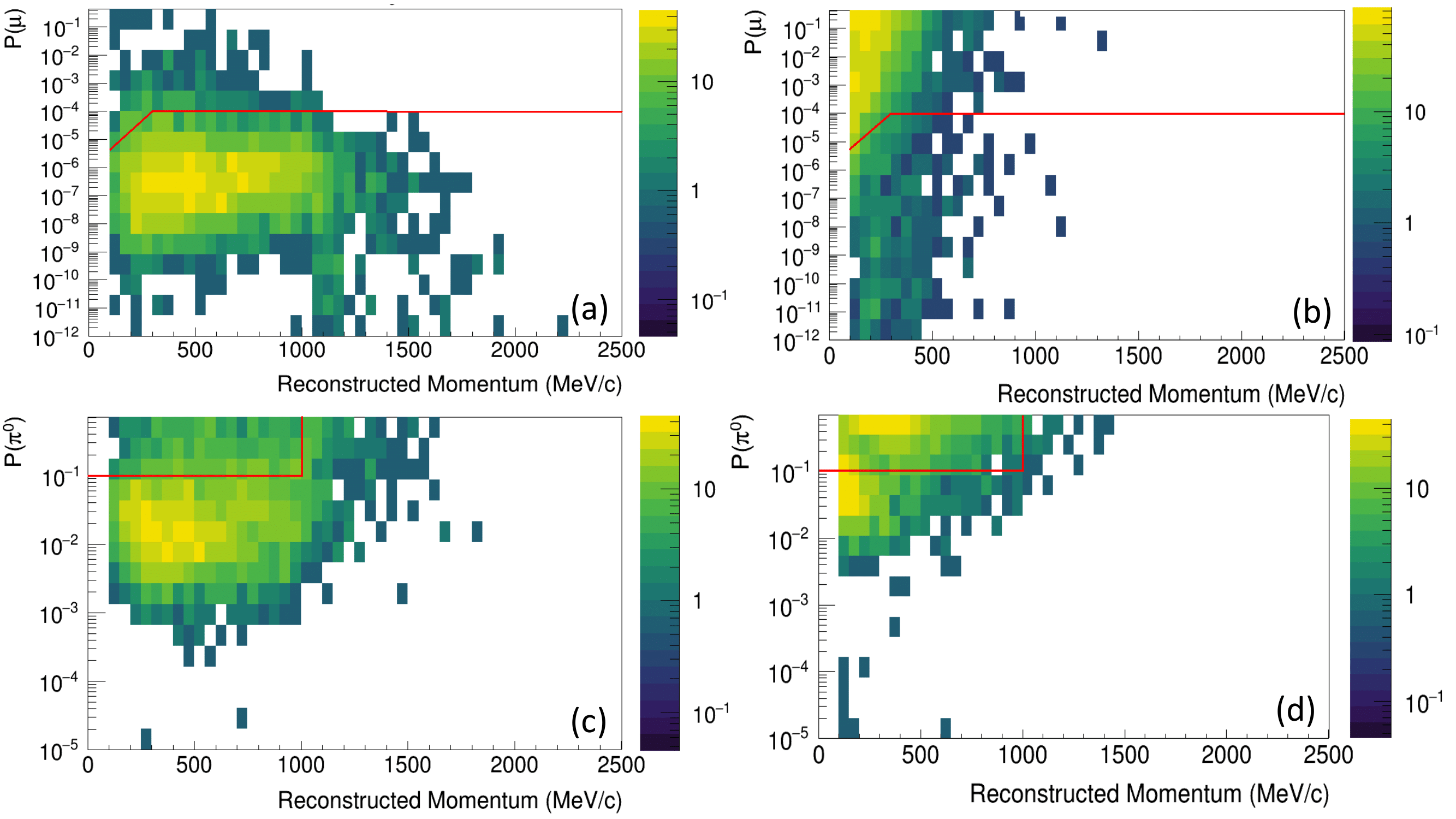}
        \caption{ 2D histograms depicting the distribution of $\nu_{e}CC0\pi$ signal in panels (a) and (c), background events $\nu_{\mu}CC$ in panel (b), and NC$\pi^{0}$ in panel (d). In each panel the softmax probability is shown with respect to the reconstructed lepton momentum.} 
        \label{ML_cuts}
    \end{minipage}
    \hfill
    \begin{minipage}[b]{0.40\linewidth}
        \centering
        \includegraphics[width=\linewidth]{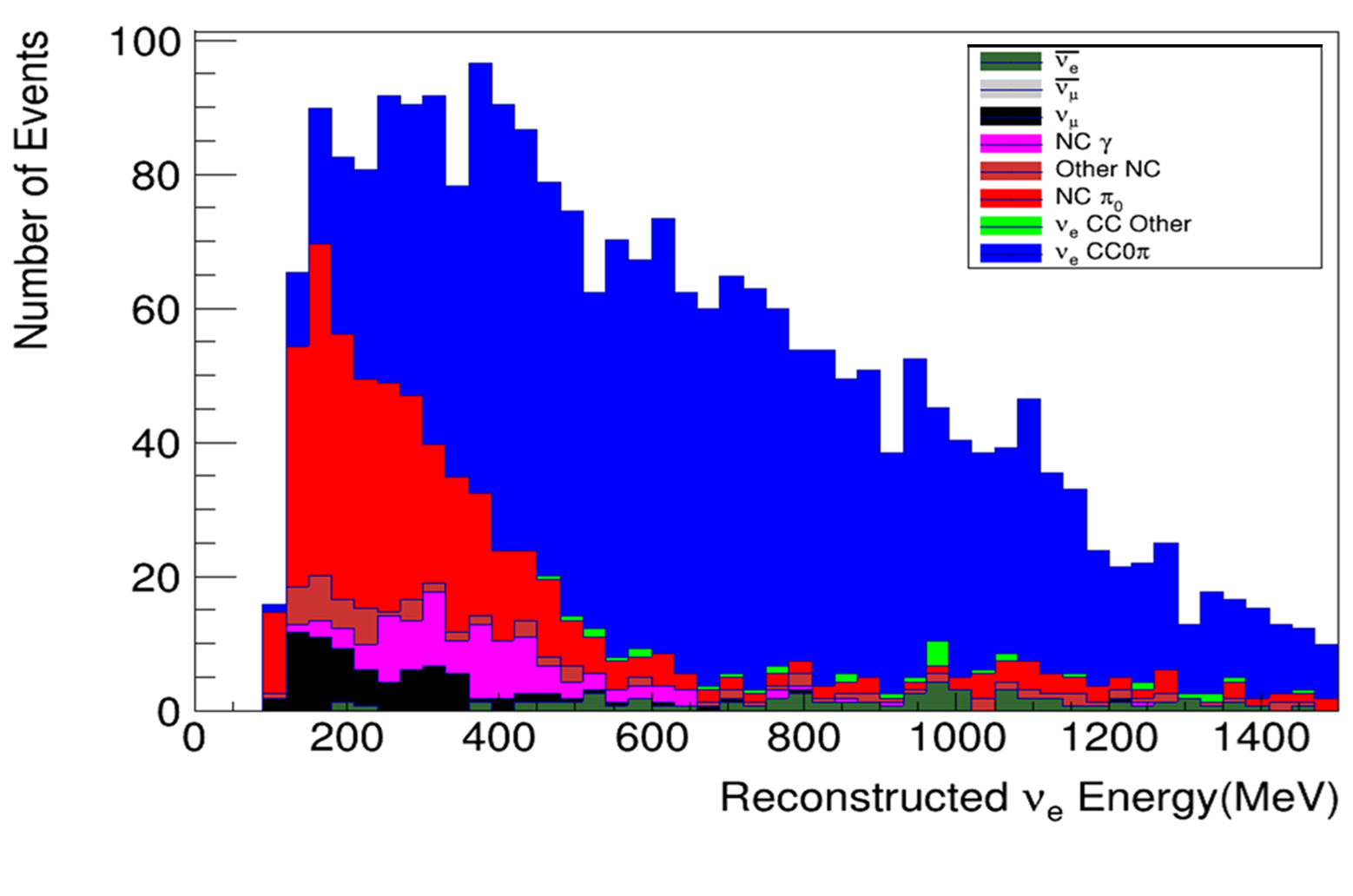}
        \caption{Interaction modes in the final selected $\nu_{e}$ events using ResNet-18 model classification.} 
        \label{all_events_ML} 
    \end{minipage}
    
\end{figure}

\vspace{-0.4cm}
Motivated by the performance of the ML framework in distinguishing particle gun events ($e^{-}$, $\mu^{-}$, $\gamma$, and $\pi^{0}$), we applied it to select $\nu_{e}$ events from a simulated beam profile. This IWCD simulated dataset contains single-ring CCQE ($\nu_{e}CC0\pi$, $\nu_{\mu}CC$), NC (NC $\gamma$, NC$\pi^{0}$), and other events such as more complex CC events with pions and/or multiple other outgoing particles in the final state with uniform energies ranging from 0 to 1.2 GeV. This test dataset is used for ML model evaluation and results in four softmax probabilities $P(e)$, $P(\mu)$, $P(\gamma)$ and $P(\pi^{0})$ corresponding to $e-$, $\mu-$, $\gamma$, and $\pi^{0}$ respectively. Then we further generated 2D histograms plotting $P(\mu)$ and $P(\pi^{0})$ softmax probability against reconstructed momentum for $\nu_{e}CC0\pi$, $\nu_{\mu}CC$ and $NC\pi^{0}$ (Figure \ref{ML_cuts}). Based on the distribution of signal and background in the histograms, we manually tuned three discriminators across $P(\mu)$, $P(\pi^{0})$, and $P(e)$ to improve both the purity and efficiency. After applying all these ML cuts sequentially along with the basic FV and kinematic variable cuts (as defined for the fiTQun-based analysis), the sample’s purity improved to 61.5\%, with an increased efficiency of 78.2\% (Table \ref{Comparison_table}). Figure \ref{all_events_ML}, showcases that by suppressing $\nu_{\mu}CC$, all ML cuts significantly reduce the number of NC events, which dominated in the fiTQun scenario previously, and hence, in this case, $\nu_{e}CC0\pi$ signal dominates over all other backgrounds (see Table \ref{Comparison_table}).

\begin{table}[ht!] 
\footnotesize
\centering
\caption{Comparison of purity and efficiency between fiTQun and ML methods for the selected $\nu_{e}$ sample separated by true interaction modes.}
\begin{tabular}{|c|c|c|c|c|c|c|c|c|c|c|}
\hline
\begin{tabular}[c]{@{}c@{}}Cuts\\ Applied\end{tabular} & $\nu_{e}CC0\pi$ & $\nu_{\mu}CC$ & \begin{tabular}[c]{@{}c@{}}Total \\ NC\end{tabular} & NC$\pi^{0}$ & NC$\gamma$ & \begin{tabular}[c]{@{}c@{}}$\nu_{e}CC$ \\ other\end{tabular} & $\bar{\nu}_{e}$ & $\bar{\nu}_{\mu}$ & Purity    & Efficiency \\ \hline
fiTQun                                                 & 2535        & 144           & 2100                                                & 1544        & 121        & 24                                                           & 157             & 1                 & 51.1$\%$ & 69.5$\%$  \\ \hline
ML                                                     & 2856        & 77            & 1342                                                & 627         & 93         & 79                                                           & 288             & 0                 & 61.5$\%$ & 78.2$\%$  \\ \hline
ML/fiTQun                                              & 1.13        & 0.54          & 0.64                                                & 0.41        & 0.77       & 3.29                                                         & 1.83            & 0                 & ---       & ---        \\ \hline
\end{tabular}
\label{Comparison_table}
\end{table} \normalsize

\vspace{-0.4cm}

\section{Conclusion}
\vspace{-0.3cm}

This work evaluated the performance of ML techniques for electron neutrino event selection in IWCD compared to fiTQun. While the latter achieved a purity of 51.1\% with an efficiency of 69.5\%, ML softmax cuts improved purity to 61.5\% with an increased efficiency of 78.2\%. With further development of more complex and automated ML cuts, we expected even more improvements beyond what is reported here.

\vspace{-0.4cm}
\section*{Acknowledgements}
\vspace{-0.2cm}

T. Mondal acknowledges the support of the Prime Minister's Research Fellowship (\href{https://www.pmrf.in/}{PMRF}). T. Mondal thanks the Mitacs Globalink Research Award (GRA) for supporting her research visit to TRIUMF, Canada.

\bibliographystyle{unsrt}
\bibliography{main_hk_ref}

\end{document}